\documentclass{eptcs}

\newif\ifereader

\ereaderfalse

\ifereader
\newcommand{\ereaderWidth}{4.20in}
\newcommand{\ereaderLength}{5.42in}

\usepackage
    [ pdftex,
      papersize={\ereaderWidth,\ereaderLength},
      twoside,
      bindingoffset=0pt,
      top=-0.3ex,bottom=1ex,
      left=0pt,right=0pt
    ]{geometry}

\fi

\providecommand{\event}{QAPL 2010} 
\providecommand{\volume}{??}
\providecommand{\anno}{2010}
\providecommand{\firstpage}{1}
\providecommand{\eid}{??}

\usepackage{ifthen}
\usepackage{stmaryrd}
\usepackage{amssymb} 
\usepackage{amsthm} 
\usepackage{amsmath} 
\usepackage{graphicx}

\DeclareMathAlphabet{\mathcal}{OMS}{cmsy}{m}{n}

\usepackage[Euler]{upgreek}

\theoremstyle{plain}
\newtheorem{theorem}{Theorem}[section]
\newtheorem{lemma}[theorem]{Lemma}

\theoremstyle{definition}
\newtheorem{definition}[theorem]{Definition}

\newcommand{\Nat}{\ensuremath{\mathbb N}}
\newcommand{\Real}{\ensuremath{\mathbb R}}
\newcommand{\RealPos}{\ensuremath{\Real_\geq}}

\newcommand{\lit}[1]{\ensuremath{\mathit{#1}}}

\newcommand{\Set}[2][]
    { \ensuremath{\ifthenelse{\equal{#1}{}}{\lit{#2}}{\lit{#2_{#1}}} }}

\newcommand{\Vars}{X}
\newcommand{\Prg}{P}
\newcommand{\Init}{\mbox{\small\ensuremath{\mathcal{I}}}}
\newcommand{\CGs}{C}
\newcommand{\Tau}{\mbox{\small\ensuremath{\mathcal{T}}}\!}

\newcommand{\totype}{\ensuremath{\ra}}

\newcommand{\CS}{\lit{\mathbb{C}S}}
\newcommand{\HS}{\lit{\mathbb{H}S}}
\newcommand{\TS}{\lit{\mathbb{T}S}}
\newcommand{\ES}{\lit{\mathbb{E}S}}
\newcommand{\GenTau}{\mathbb{G}\Tau}
\newcommand{\Power}{\lit{\mathbb{P}}} 
\newcommand{\PowerNonEmpty}{\lit{\mathbb{P}^+}}
\newcommand{\Def}{\ensuremath{\triangleq}}
\newcommand{\Subst}[2][]
	{ \ensuremath{\ifthenelse{\equal{#1}{}}{\langle #2 \rangle}{\langle #1\mapsto #2 \rangle} }}

\newcommand{\ra}{\ensuremath{\rightarrow}}

\newcommand{\dotsep}{\mbox{\raisebox{0.4ex}{\tiny\ensuremath{\:\;\bullet\;\:}}}}

\newcommand{\Eq}{\ensuremath{\mathop{\:\equiv\:}}}
\renewcommand{\And}{\ensuremath{\mathop{\:\wedge\;}}}

\newcommand{\Impl}{\ensuremath{\Rightarrow}}
\newcommand{\Not}{\ensuremath{\neg}}

\newcommand{\Const}[1]{\ensuremath{\underline{#1}}} 
\newcommand{\SubPr}[1]{\ensuremath{\overline{#1}}} 

\newcommand{\Wp}{\lit{wp}}

\newcommand{\Min}{\ensuremath{\sqcap}}

\newcommand{\SKIP}[0]{\textbf{skip}}
\newcommand{\ABORT}[0]{\textbf{abort}}
\newcommand{\IF}[0]{\textbf{if}}
\newcommand{\THEN}[0]{\textbf{then}}
\newcommand{\ELSE}[0]{\textbf{else}}
\newcommand{\FI}[0]{\textbf{fi}}
\newcommand{\DO}[0]{\textbf{do}}
\newcommand{\OD}[0]{\textbf{od}}

\newcommand{\True}[0]{\textsf{true}}
\newcommand{\False}[0]{\textsf{false}}

\newcommand{\slq}{\ensuremath{\sqsubseteq}}
\newcommand{\slqn}{\subseteq}

\newcommand{\elq}{\ensuremath{\Rrightarrow}}

\newcommand{\aelq}{\ensuremath{\Rrightarrow^\sharp}}

\let\oldmarginpar\marginpar
\renewcommand\marginpar[1]{\-\oldmarginpar[\raggedleft\footnotesize #1]%
{\raggedright\footnotesize #1}}
\newcommand{\remark}[2]{}
\newcommand{\remarkN}[1]{\remark{N}{#1}}
\newcommand{\remarkD}[1]{\remark{D}{#1}}

\newcommand{\noi}[0]{\noindent}

\title{
  Automatic Probabilistic Program Verification through Random Variable Abstraction
}

\author{
  Dami\'an Barsotti
  \thanks{Granted by MinCyT PID2008, Gobierno de la Provincia de C\'ordoba.}
\and
  Nicol\'as Wolovick
  \footnotemark[1]
\end{tabular}\\[2ex]
  \footnotesize{Fa.M.A.F., Universidad Nacional de C\'ordoba, Ciudad Universitaria, 5000
  C\'ordoba, Argentina}
\begin{tabular}[t]{c}
}

\def\titlerunning{Automatic Probabilistic Program Verification through Random Variable Abstraction}
\def\authorrunning{D. Barsotti \& N. Wolovick}

\begin{document}
\maketitle

\begin{abstract}
  The weakest pre-expectation calculus~\cite{morgan04arp} has been
  proved to be a mature theory to analyze quantitative properties of
  probabilistic and nondeterministic programs. We present an automatic
  method for proving quantitative linear properties on any denumerable
  state space using iterative backwards fixed point calculation in the
  general framework of abstract interpretation. In order to accomplish
  this task we present the technique of random variable abstraction
  (RVA) and we also postulate a sufficient condition to achieve
  exact fixed point computation in the abstract domain.  The
  feasibility of our approach is shown with two examples, one
  obtaining the expected running time of a probabilistic program, and
  the other the expected gain of a gambling strategy.

  Our method works on general guarded probabilistic and
  nondeterministic transition systems instead of plain pGCL programs,
  allowing us to easily model a wide range of systems including
  distributed ones and unstructured programs. We present the
  operational and $\Wp$ semantics for this programs and
  prove its equivalence.

\end{abstract}



\section{Introduction}\label{intro}

Automatic probabilistic program verification has been a field of active research in the last two decades.
The two major approaches to tackle this problem have been model checking~\cite{vardi85auto,hansson94pctl,bianco95mcp,hinton06prism}, and theorem proving~\cite{hurd05mech,celiku06mech}.
Traditionally model checking has been targeted to be a push-button technique, but in the quest of full automation some restrictions apply.
The most prominent one is finiteness of the state space, that leads to finite-state Markov chains (MC) and Markov decision processes (MDP).
These models are usually verified against probabilistic extensions of temporal logics such as PCTL~\cite{hansson94pctl,bianco95mcp}.
Even with the finiteness restriction, the state explosion problem have to be alleviated using, for example, partial order reduction techniques~\cite{baierDG06por}.
Another approach is taken in the PASS tool~\cite{wachterZH07pass,hermannsWZ08cegar}, where the authors profit from the work on predicate abstraction~\cite{graf97pa,uribe98abstractions} in the general framework of abstract interpretation~\cite{cousot77ai} in order to aggregate states conveniently and prevent the state explosion problem or even handle infinite state spaces.
Theorem proving techniques can also overcome state explosion and infinite state space problems but at a cost, the automation is up-to loop invariants, that is, once the correct loop invariants are fixed, the theorem prover can automatically prove the desired property~\cite{hurd05mech}.
There are also mixed approaches like~\cite{gordo08refutation} in the realm of refinement checking of pGCL programs.

Automatic probabilistic program verification cannot, to the author's knowledge, tackle the problem of performance measurement of possibly unbounded program variables.
Typical examples of this quantitative measurements are expected number of rounds of a probabilistic algorithm, or the expected revenue of a gambling strategy.
In this kind of problems, the model checking approach discretize the variable up to a bound, either statically or using counterexample-guided abstraction refinement (CEGAR)~\cite{hermannsWZ08cegar}, but since the values it can reach are unbounded, an approximation is needed at some point.
In theorem proving this problem could be solved but the invariants have to devised, and anyone involved in this task knows that this is far from trivial.

We propose the computation of parametric linear invariants on a set of predefined predicates, and we call this random variable abstraction (RVA) as it generalizes predicate abstraction (PA) to a quantitative logic based on expectations~\cite{morgan04arp}.
Our motivation is simple.
If the underlying invariants of the program can be captured as a sum of linear random variables in disjoint regions of the state space, a suitable fixed point computation should be able to discover the coefficients of such a linear expressions.
To compute the coefficients we depart from the traditional forward semantics approach, and use weakest \mbox{pre-expectation} ($\Wp$) backwards semantics~\cite{morgan04arp}.
This combination of predicate abstraction, linear random variables and quantitative $\Wp$, renders a new technique that can compute, for example, expected running time of probabilistic algorithms, even though this quantities are not a-priori bound by any constant.
This allows us to reason about efficiency of algorithms.

Parametric linear invariants for probabilistic and nondeterministic programs using expectation logic are also generated in~\cite{katoen09linear}, but our work differs in two aspects.
First, in~\cite{katoen09linear} all random variables have to be one-bounded or equivalently they should be bounded by a constant.
Second, in the computation technique for the coefficients, they cast the problem to one of constraint-solving and resort to off-the-shelf constraint solvers, while we use a fixed point computation.

\smallskip

\textit{Outline.} In Section~\ref{motivation} we first give a motivating example that briefly shows the type of problems we address as well as the mechanics to obtain the result.
Then in Section~\ref{concrete_semantics} the probabilistic and nondeterministic programs, its operational and $\Wp$ semantics are presented, as well as the fixed point computation in the concrete domain of (general) random variables.
Section~\ref{abstract_semantics} covers the abstract domain of random variables, its semantics, and the fixed point computation, as well as the general problems to face in this process.
A more involved example is given in Section~\ref{case_study}.
Section~\ref{conclusions} concludes the paper and discusses future work.

\section{Motivating Example}\label{motivation}

In order to give some intuition on what we are going to develop throughout the paper, we present a purely probabilistic program $\Prg_0$ that generates a geometric distribution.
\begin{figure}[!ht]
\begin{center}
\begin{minipage}{\textwidth}
\begin{tabbing}
	$\Prg_0$\ $\Def$ \= \ \= $x,i:=1,0$ \\
	\>; \> $\DO$ \= $x\neq 0 \ra$ \\
	\>\>	\> \ \= $x:=0 \oplus_{\frac{1}{2}} x:=1$ \\
	\>\>	\> ; \> $i:=i+1$ \\
	\>\> $\OD$ 
\end{tabbing}
\end{minipage}
\end{center}
\caption{Geometric distribution.}\label{fig:ex_geomdist}
\end{figure}
This program halts with probability 1 and variable $i$ could reach any natural number with positive probability.
The semantics of this program can be regarded as a probability measure $\Delta$ over the naturals with distribution $\Delta.\{ K<i\} = \tfrac{1}{2^K}$.
The expected running time of the algorithm is precisely the expectation of random variable $i$ with respect to the measure $\Delta$, that is $\int_\Delta i$.
Informally we could do this quantitative analysis as follows:
the first iteration is certain, however the second one occurs with probability one half, the third one will occur with half the probability of previous one, and so on.
In~\cite[p.69]{morgan04arp} the same calculation is done by hand, using the invariant $(i+K) \textit{\ if\ } (x\neq 0) \textit{\ else\ } i$, the equations fix $K=2$ and this implies that given the initialization, the expected value of random variable $i$ is $2$.

Our approach establishes a recursive higher-order function $f$ on the domain of random variables using weakest pre-expectation semantics~\cite{morgan04arp}.
It computes the expectation of taking a loop cycle and repeating, or ending.
\[ f.X \Def [x\neq 0]\times \Wp.(x:=0 \oplus_{\frac{1}{2}} x:=1; i:=i+1).X + [x=0]\times i \]
We are looking for a \emph{fixed point} of this functional, namely $f.\varphi=\varphi$, but taking a specific subset of random variables: two linear functions on the post-expectation $i$, one for each region defined by the predicates that control the flow of the program.
It is a \emph{parametric random variable} with coefficients $((a_1,a_0),(b_1,b_0))$.
\[ \textit{Inv} \Def (a_1 i + a_0) [x=0] + (b_1 i + b_0) [x\neq0] \]
Calculating $f$ in the random variable $\textit{Inv}$:
\[ f.\textit{Inv} = (1 i + 0) [x=0] + (\tfrac{a_1+b_1}{2}i + \tfrac{a_0+a_1+b_0+b_1}{2}) [x\neq0]\enspace. \]
%
If we only look at the coefficients, $f$ can be (exactly) captured by function $f^\sharp$ on the domain $\Real^{(1+1)\times 2}$.
\[ f^\sharp.((a_1,a_0),(b_1,b_0)) = ((1,0),(\tfrac{a_1+b_1}{2},\tfrac{a_0+a_1+b_0+b_1}{2})) \enspace.\]
The computation goes as follows, from the random variable that is constantly $0$ represented by the tuple $((0,0),(0,0))$ we apply the update rule iteratively until the floating point precision produces a fixed point after a little less than half hundred iterations (Table~\ref{table:geometric}).
\begin{table}[!ht]
\begin{small}
\begin{center}
\begin{tabular}{|c||c|l|}
\hline
& \multicolumn{2}{|c|}{Coefficients} \\
\hline
Iter. & ($a_1$, $a_0$) & ($b_1$, $b_0$)  \\
\hline
0  &  (0, 0) & (0, 0) \\
1  &  (1, 0) & (0, 0) \\
2  &  (1, 0) & (0.5, 0.5) \\
3  &  (1, 0) & (0.75, 1) \\
4  &  (1, 0) & (0.875, 1.375) \\
5  &  (1, 0) & (0.9375, 1.625) \\
6  &  (1, 0) & (0.96875, 1.78125) \\
7  &  (1, 0) & (0.984375, 1.875) \\
\vdots & \vdots & \phantom{(1,}\vdots  \\
45  &  (1 ,0) & (1, 2)\\
\hline
\end{tabular}
\end{center}
\end{small}
\caption{Iteration for the geometric distribution program.}
\label{table:geometric}
\end{table}

These fixed point coefficients represent the random variable $i\times [x=0] + (i + 2) \times [x\neq0]$ that is a valid \mbox{pre-expectation} of the repeating construct with respect to post-expectation $i$.
If we take the weakest \mbox{pre-expectation} of initialization (syntactic substitution) we obtain the value $2$.
Note that invariant and expectation of random variable $i$ coincide with the calculations done by hand in~\cite{morgan04arp}.


\section{Concrete Semantics}\label{concrete_semantics}

\subsection{Probabilistic Programs}

We fix a finite set of \emph{variables} $\Vars$. 
A \emph{state} is a function from variables to the semantic domain $s:\Vars\ra \Omega$, and the set of all states is denoted by $S$.
An \emph{expression} is a function from states to the semantic domain $\beta: S \ra \Omega$, while a \emph{boolean expression} is a function $G:S\ra \{\True,\False\}$, and an \emph{assignment} $E:S\ra S$ is a state transformer.
We define \emph{substitution} of expression $\beta$ with assignment $E$, $\beta\Subst{E}$ as function composition $\beta\circ E$.

A \emph{guarded command} $(G \ra E_1@p_1 | \cdots | E_k@p_k)$ consists of a boolean guard $G$ and assignments $E_1,\cdots, E_k$ weighted with probabilities $p_1,\cdots, p_k$ where $\sum_{i=1}^k p_i \leq 1$.

A \emph{program} $\Prg = (S, \Init, \CGs)$ consists of a boolean expression $\Init$ that defines the set of initial states and a finite set of guarded commands $\CGs$.

Note that using this guarded commands we can underspecify (in the sense of refinement) in two ways.
One way is using guard overlapping, so in the non-null intersection of $G_0$ and $G_1$ there is a nondeterministic choice between the two multi-way probabilistic assignments.
The other is using subprobability distributions, since the expression $(E_0@\tfrac{1}{4} | E_1@\tfrac{1}{4})$ can be regarded as all the probability distributions that assign at least $\tfrac{1}{4}$ to each branch~\cite{josee09demonic}.
Later, in the operational model, we will see that the first one is handled by convex combination closure, while the later is handled by up closure.
\remarkD{Se podria dibujar el ejemplo primero como transys}

\subsection{Operational model}

We fix the semantic domain $\Omega$ as a denumerable set, for example the rationals.
\remarkN{explicar porque, asunto de medibilidad de $\Tau$}
The power set is denoted by $\Power$, and if the empty set is not included we denote it $\PowerNonEmpty$.

The set of (discrete) \emph{sub-probability measures} over $S$ is
$\SubPr{S} = \{ \Delta: S\ra [0..1] \mid \sum\Delta\leq 1 \}$.
The partial order $\Delta\sqsubseteq\Delta'$ on $\SubPr{S}$ is defined pointwise, and $\Const{0}$ denotes the least element everywhere defined $0$.
A set of sub-probability measures $\xi\subseteq\SubPr{S}$ is \emph{up-closed} if $\Delta\in\xi$ and $\Delta\sqsubseteq\Delta'$ then $\Delta'\in\xi$.
Similarly $\xi$ is \emph{convex closed} if for all $p\in[0..1]$ and $\Delta_0,\Delta_1\in\xi$ implies $p\Delta_0+(1-p)\Delta_1\in\xi$.
Finally $\xi$ is \emph{Cauchy-closed} or simply closed if it contains its boundary, that is for every sequence $\{\Delta_i\}\subseteq\xi$, such that $\Delta_i \stackrel{i\ra\infty}{\longrightarrow} \Delta$, then $\Delta\in\xi$.
\remarkN{no me gusta como queda el limite}
$\CS$ is the set of \mbox{non-empty}, \mbox{up-closed}, convex and \mbox{Cauchy-closed} subsets of $\SubPr{S}$.
%
%
The set of probabilistic programs over $S$ is defined $\HS \Def S \ra \CS$, and it is ordered pointwise with ${}\sqsubseteq{}$.

The \emph{probabilistic and nondeterministic transition system} $\mathcal{M}$ is a tuple $(S, \Init, \Tau)$
where $S$ is the set of states, $\Init \subseteq S$ is a set of initial states, and an $\Tau\in \HS$ is called \emph{transition function}.


The semantics of a program $\Prg = (S, \Init, \CGs)$ is the tuple $\mathcal{M} = (S, \Init, \Tau)$ with the same state space and initial states, where $\Tau.s$ is defined by its generators.
\begin{definition}
	Given the program $\Prg = (S, \Init, \CGs)$,
	where $\CGs=\{ i:[0..l) \dotsep (G_i \ra  E^i_1 @ p^i_1 | \cdots | E^i_{k_i} @ p^i_{k_i} ) \}$,
	the \emph{set of generators of $\Tau.s$} is 
	$\GenTau.s \Def \{ i:[0..l) \mid G_i.s \dotsep \Delta_i \} $ if $(\exists i:[0..l) \dotsep G_i.s)$,
	otherwise $\GenTau.s=\{\Const{0}\}$,
	where $\Delta_i.s' = ( \sum j: [1..k_i] \mid E_j^i.s = s' \dotsep p_j^i )$\enspace.
\end{definition}
The set $\GenTau.s$ is nonempty and closed. 
Now $\Tau.s$ can be defined from above as the minimum set over $\CS$ that includes $\GenTau.s$.
Note this set is well defined since $\GenTau.s\subseteq\SubPr{S}\in \CS$, and $\CS$ is closed by arbitrary intersections. 
We can also define $\Tau.s$ from below, taking the up and convex closure of the generator set $\GenTau.s$.
There is no need to take the Cauchy closure since the generator set is closed and the operators that form the up and convex closure preserve this property.

\subsection{Expectation-transformer semantics} \label{sec:expct-trans_semantic}

In the previous section a probabilistic and nondeterministic program semantics was defined as a transition function from states to a subset of subprobability distributions.
This set was defined to be up and convex closed.

Verification of this particular type of programs poses the question of what propositions are in this context.
We take the approach by Kozen~\cite{kozen83pdl}, where propositions $Q: S\ra \{\True,\False\}$ are generalized as \emph{random variables}, that is a measurable function $\beta:S\ra \RealPos$ from states to positive reals.
The program $P$ is taken as a distribution function $\Delta$, and the program verification problem, formerly a boolean value $P\models Q$ is cast into an integral $\int_\Delta \beta$ that is the \emph{expectation} of random variable $\beta$ given distribution $\Delta$.
In a nutshell this is a quantitative logic based on expectations.
Boolean operators are (no uniquely) generalized.
The arithmetic counterpart of the implication is $\alpha\elq\beta$, that is pointwise inequality $(\forall s:S \dotsep \alpha.s\leq\beta.s)$.
Conjunction ${}\land{}$ can be taken as multiplication $\times$ or minimum ${}\sqcap{}$, and disjunction $\lor$ over non overlapping predicates is captured by addition ${}+{}$.
The characteristic function operator $[\cdot]$ goes from the boolean to the arithmetic domain with the usual interpretation $[\True]=1$, $[\False]=0$, and using it we could define negation as subtracting the truth value from one $[\Not Q]=1-[Q]$.

This model generalizes the deterministic one since it is proven that $[P\models Q] = \int_P [Q]$ for predicate $Q$ and deterministic program $P$.

In~\cite{he97models} He et al. extend Kozen's work to deal with nondeterminism as well as probabilism in the programs.
The program verification problem is again a real value, namely the least expected value of the random variable with respect to the sub-probability distributions generated by the program (demonic nondeterminism).

Probabilistic (and nondeterministic) program verification problem is usually defined in the program text itself as Hoare triple semantics or the equivalent weakest precondition calculus of Dijkstra~\cite{kozen83pdl,morgan04arp}.
The triple $\{Q\}P\{Q'\} \equiv Q\Rightarrow \Wp.P.Q'$ is generalized to $\{\alpha\}P\{\beta\} \equiv \alpha\elq\Wp.P.\beta$, and the generalized $\Wp$ is called \emph{weakest pre-expectation calculus} or probabilistic $\Wp$ semantics.

In general, the program outcome depends on the input state $s$, therefore $\Wp.P.\beta$ is a function that given the initial state establish a lower bound on the expectation of random variable $\beta$ (it is exact in the case of a purely probabilistic program).
Therefore, expectations as functions of the input are also random variables, hence we define the \emph{expectation space} as $\ES \Def \langle S\ra\RealPos, \elq\rangle$.
It is worth noticing that all functions in $\ES$ are measurable since $S$ is taken to be denumerable (all subsets of $S$ are measurable).
This implies that is valid to write integrals like $\int_\Delta f$ with $\Delta\in\SubPr{S}$ and $f\in\ES$.
We define the \emph{expectation transformer space} as the functions from expectations to expectations $\TS \Def \langle \ES\leftarrow \ES, \sqsubseteq\rangle$.

The $\Wp$ expectation calculus is defined structurally in the constructors of probabilistic and nondeterministic programs (Fig.~\ref{fig:wp_calc}).
This function basically maps assignments to substitutions, choice and probabilistic choice to convex combination and nondeterministic choice to minimum. 
\begin{figure}[!ht]
\begin{small}
\begin{align*}
	\Wp.\ABORT.\beta &\Def \Const{0} \\
	\Wp.\SKIP.\beta &\Def \beta \\
	\Wp.E.\beta &\Def \beta\Subst{E} \\
	\Wp.(P_0;P_1).\beta &\Def \Wp.P_0.(\Wp.P_1.\beta) \\
	\Wp.(P_0 \oplus_p P_1).\beta &\Def p\times\Wp.P_0.\beta + (1-p)\times\Wp.P_1.\beta \\ 
	\Wp.(\IF\ G\ \THEN\ P_0\ \ELSE\ P_1\ \FI).\beta &\Def \Wp.(P_0\oplus_{[G]}P_1).\beta \\ 
	\Wp.(P_0 \Min P_1).\beta &\Def \Wp.P_0.\beta \Min \Wp.P_1.\beta \\ 
	\Wp.(\DO\ G\ra P\ \OD).\beta &\Def (\mu X \dotsep [G]\times \Wp.P.X + [\Not G] \times \beta)
\end{align*}
\end{small}
\caption{$\Wp$ expectation calculus.}
\label{fig:wp_calc}
\end{figure}

\sloppy Inspired in this semantics, we are going to define a weakest pre-expectation of our program \mbox{$\Prg = (S, \Init, \CGs)$}.
We first transform the program $\CGs$ into a semantically equivalent set but with pairwise disjoint guards.
Let $\CGs = \{ i:[0..l) \dotsep (G_i \ra  E^i_1 @ p^i_1 | \cdots | E^i_{k_i} @ p^i_{k_i} ) \}$.
We follow the standard approach of taking the complete boolean algebra~\cite{IntroductionLatticeOrderBook} over the finite set of assertions $\{ G_i \}$.
The new program $\Prg'$ is:
\[ \Prg' = \{ I:\PowerNonEmpty{[0..l)}, i:I \dotsep (A_I \ra  E^i_1 @ p^i_1 | \cdots | E^i_{k_i} @ p^i_{k_i} ) \} \]
where
\begin{equation} \label{eq:guard_atoms} 
A_I = (\land i:I \dotsep G_i) \land (\land i:[0..l)\!-\!I \dotsep \neg G_i)\enspace.
\end{equation}
This program can be regarded as a pGCL program that consists of an if-else ladder of the atoms $A_I$.
In each branch $I$, there is a $|I|$-way nondeterministic choice of $k_i$-way probabilistic choice with $i\in I$.
\remarkN{notar que no puedo hacer sub-prob con los constructores estandar}
The $\Wp$ semantics is defined as follows:
\begin{equation}
\label{eq:wp_sint}
\Wp{}.\Prg'.\beta \Def \left( \sum I:\PowerNonEmpty[0..l) \dotsep
	[A_I] \times \left( \bigsqcap i:I \dotsep
				\left( \sum j:[1..k_i] \dotsep p_i^j\times \beta\Subst{E_i^j} \right)
			  \right)
\right)\enspace.
\end{equation}

\subsection{Relational to expectation-transformer embedding}

We now define the notion of $\Wp$ but in terms of the transition $\Tau$. 
The injection $\Wp \in \HS \totype \TS$ is defined as the minimum expectation for random variable $\beta$ over all nondeterministic choices of sub-probability distributions.
\begin{equation}
\label{eq:wp_sem}\textstyle
	\Wp{}.\Tau.\beta.s \Def \left(\bigsqcap \Delta : \Tau.s \dotsep {\textstyle{\displaystyle\int}_{\!\!\!\Delta}} \beta\right)\enspace.
\end{equation}
%
The next lemma shows that the syntactic definition (\ref{eq:wp_sint}) inspired in pGCL $\Wp$ calculus coincides with the semantic one (\ref{eq:wp_sem}).
From now on we will use both definitions interchangeably.
\begin{lemma}
	The syntactic and semantic notions of weakest pre-expectation coincide, that is
	$\Wp{}.\Prg'.\beta.s = \Wp{}.\Tau.\beta.s$\enspace.
\end{lemma}
\begin{proof}
First suppose there is a valid $G_i.s$.
We begin by the rhs of the equality, that by (\ref{eq:wp_sem}) is $(\Min \Delta : \Tau.s \dotsep {\textstyle{\int}_{\!\!\!\Delta}} \beta)$.
The equality will be first proved for the generators $\GenTau.s\subseteq\Tau.s$,
where $\GenTau.s =$ \mbox{$\{ i:[0..l) \mid G_i.s \dotsep \Delta_i \}$}
and $\Delta_i.s' = ( \sum j: [1..k_i] \mid E_j^i.s = s' \dotsep p_j^i )$.
Let $I = \{i:[0..l) \mid G_i.s\}$ the index set of valid guards, then the minimum over all expectations is
$(\Min i:I \dotsep {\textstyle{\int}_{\!\!\!\Delta_i}} \beta) $.
Now we develop the inner term using the fact that $\Delta_i.s'\neq 0$ on a finite set of points.
\begin{align*}
	{\textstyle{\int}_{\!\!\!\Delta_i}} \beta & = (\textstyle{\sum} s':S \dotsep \beta.s'\times \Delta_i.s') & \\
	& =  ( \textstyle{\sum} s':S \dotsep \beta.s'\times (\sum j: [1..k_i] \mid E_j^i.s = s' \dotsep p_j^i) ) & \mbox{\{ definition $\Delta_i$ \}}\\
	& =  ( \textstyle{\sum} s':S, j: [1..k_i] \mid E_j^i.s = s' \dotsep \beta.s' \times p_j^i ) & \mbox{\{ distributivity \}} \\
	& =  ( \textstyle{\sum} j: [1..k_i] \dotsep p_j^i \times \beta.(E_j^i.s) ) & \mbox{\{ $s'$ is a function of $j$ \}} \\
	& =  ( \textstyle{\sum} j: [1..k_i] \dotsep p_j^i \times (\beta\Subst{E_j^i}).s ) & \mbox{\{ definition of substitution \}} \\
\end{align*}
Summing up we have
\[ \Wp{}.\Tau.\beta.s = \left( \Min i:I \dotsep \left(\sum j: [1..k_i] \dotsep p_j^i \times (\beta\Subst{E_j^i}\right).s\right)\enspace. \]
Taking the expression $\Wp{}.\Prg'.\beta$ in (\ref{eq:wp_sint}) and evaluating in $s$ we get the same expression since only $A_I$ is valid.

The other case is given when all guards are invalid in $s$.
The rhs of the equation is $0$ as $\Const{0}\in\Tau.s$, and the lhs is a finite sum of $0$'s.

For the rest of subprobability measures generated by the up and convex closure in $\GenTau.s$ we show the minimum is maintained.
Let $\Delta_0$ be the measure achieving the minimum expectation.
If for up-closure it was added $\Delta\sqsupseteq \Delta_0$, by monotonicity of the integral we would get a greater expectation ${\textstyle{\int}_{\!\!\!\Delta}} \beta \geq {\textstyle{\int}_{\!\!\!\Delta_0}} \beta$.
For and added convex combination $p\Delta_0 + (1-p) \Delta_1$, it is easy to show that $\Delta_0 = p\Delta_0 + (1-p) \Delta_0\leq p\Delta_0 + (1-p) \Delta_1$, and the minimum is not changed.
\end{proof}

\subsection{Fixed points}

The theory of fixed points is frequently used to give meaning to iterative
programs like the ones we outlined in Section~\ref{sec:expct-trans_semantic}. In this
section we present a fixed point theory in accordance with our particular domain.

A poset is a \emph{\mbox{meet-semilattice}} if each pair of elements has a greatest
lower bound (it has meet). An \emph{almost complete \mbox{meet-semilattice}} is a
meet-semilattice where every
non-empty subset has a greatest lower bound (it has infimum).
\remarkD{almost complete \Impl local cpo (sale en paper). Creo qe esto
  ultimo es mas debil. Si algo no anda con las expectations se puede
  repensar con local cpo}
The poset $\langle\RealPos,\leq\rangle$ agrees this property by completeness of the real numbers.
Given a poset $\langle X, \slq \rangle$, a function $f:X \totype X$ is
\emph{monotone} if $x\slq y \Impl f.x\slq f.y$.
%
An element $x \in X$ is a \emph{pre-fixed point} if $f.x \slq x$, and if $f.x =
x$ then is a \emph{fixed point}.  The \emph{least fixed point} is denoted
as $\mu.f$ if it exists.

It's easy to see that $\ES$ is an almost complete \mbox{meet-semilattice}
because it is the pointwise extension of $\langle\RealPos,\leq\rangle$ from the set of states
$S$, then it inherits the property.
Therefore, we will develop the basic
fixed point theory focused on these domains.  The next result (proved in \cite[Lemma
  2.15]{IntroductionLatticeOrderBook}) establishes the existence of the supremum
of a subset of an almost complete \mbox{meet-semilattice}:

\begin{lemma} \label{lem:priestley}
Let $P$ be an almost complete \mbox{meet-semilattice}. Then the supremum $\sqcup
S$ exists in $P$ for every subset $S$ of $P$ which has an upper bound in $P$.
\end{lemma}

In general, an almost complete \mbox{meet-semilattice} is not necessarily a
complete partial order (CPO). For example, the poset $\langle\RealPos,\leq\rangle$ does not
form a CPO (it lacks a top element) but it is an almost complete
\mbox{meet-semilattice}. Hence, we cannot prove the general existence of fixed
points in this domain.  Although, under certain conditions, existence of fixed
points is guaranteed as stated in~\cite[Theorem 1.4]{Mislove93fixedpoints}:





\begin{theorem} \label{th:fixpoint}
  Let a poset $\langle X, \slq \rangle$ and a monotone function $f : X \totype
  X$. Suppose $X$ is an almost complete meet-semilattice which has a least
  element $\bot$ and $f$ has a \mbox{pre-fixed} point in $X$. Then $f$ has a
  least fixed point as the infimum of the set of the pre-fixed points of $f$:
  $\mu.f = (\sqcap \, x \;|\; f.x \slq x \dotsep x)$\enspace.




\end{theorem}





\noi The next section shows a suitable notion of correctness for our programs,
using this theorem.


\subsection{Correctness} \label{sec:correctness}

The $\Wp$ semantics allows us to express quantitative properties as
expectations. In order to verify this properties, we present a notion of
correctness based on the fact that any program $P = (S,\Init,C)$ can be written
as a pGCL program. Such programs can be constructed as a while loop with
body made as a nested \mbox{if-else} ladder statement as we mentioned in
Section~\ref{sec:expct-trans_semantic}.  The guards of these conditional
statements are the predicates $A_I$ in (\ref{eq:guard_atoms}) and its bodies are
the non-deterministics probabilistic assignments corresponding to each guard as
defined in~(\ref{eq:wp_sint}).
The loop guard $G$ is the disjunction of the atoms, so it exits the loop if there is no valid guard in the if-else ladder.
The weakest pre-expectation semantics of this
program is defined in \cite{morgan04arp} as the least fixed point:
\[
  \Wp.\DO\ G \ra P'\ \OD.\beta 
  = (\mu X \dotsep [G] \times \Wp.P'.X + [\Not G] \times \beta)
\]
where $G = (\vee \, I:\PowerNonEmpty[0..l) \dotsep A_I)$ and $P'$ is the
\mbox{if-else} ladder statement. Therefore, we can characterize the following
idea of correctness:


\begin{definition}
  Let a program $P = (S, \Init, C)$ with $G_0,\cdots, G_{l-1}$ the guards of the
  commands in $C$, and two expectations $\alpha, \beta \in \ES$. Let also $G \Def (\vee
  \, i : [0..l) \dotsep G_i)$ and $f : \ES \totype \ES$ the expectation
    transformer defined as:

  \begin{equation} \label{eq:back_propag_def}
    f.X \Def [G] \times \Wp.P.X + [\Not G] \times \beta \enspace.
  \end{equation}

  \noi Then, we said that $[\Not G]\times\beta$ is a valid \mbox{post-expectation} of $P$ with
  respect to the \mbox{pre-expectation} $\alpha$ if
  \begin{equation} \label{eq:initialization_vc}
    \alpha \elq \mu.f \enspace.
  \end{equation}
\end{definition}

\smallskip

Based on this definition, if we can find the fixed point $\varphi = \mu.f$
defined from (\ref{eq:back_propag_def}) then we can show that 
\begin{align*}
            [G] \times \varphi &\elq \Wp.P.\varphi\\
\text{and } [\Not G] \times \varphi &\elq \beta \enspace,
\end{align*}
and these equations denote $\varphi$ as an invariant of
the system verifying the \mbox{post-expectation} $\beta$. Also, if it
satisfies (\ref{eq:initialization_vc}), the initialization of the loop
is also fulfilled.

\smallskip
Moreover, our definition of correctness requires the existence of a least fixed
point~$\mu.f$. Although the poset $\langle \ES, \elq \rangle$ is an almost
complete \mbox{meet-semilattice}, it is not a CPO because $\RealPos$ itself is not
(it lacks an adjoined $\infty$ element). Therefore, we must remark the
existence of the least fixed point $\mu.f$ is guaranteed only if $f$ has a
\mbox{pre-fixed} point as we require in Theorem~\ref{th:fixpoint}. There are
many cases where \mbox{pre-fixed} point do not exist, even for plain
pGCL programs. An example of this kind of program is $P_0'$ where it is obtained from
Fig.~\ref{fig:ex_geomdist} simply replacing $i:=i+1$ by the (also linear)
$i:=2*i$.
With this modification, it can be calculated that $\Wp.P_0'.i = \infty$.

\section{Abstract Semantics}\label{abstract_semantics}

\subsection{Preliminaries}

Following \cite{monniaux00}, we review the framework of abstract interpretation
applied to our problem.  Let us consider two posets $\Gamma = \langle X, \slq
\rangle$ and $\Gamma^\sharp = \langle X^\sharp, \slqn \rangle$, and a monotone
function $\gamma : X^\sharp \totype X$ called \emph{concretization function}. An
element $x^\sharp \in X^\sharp$ is said to be a \emph{backward abstraction}
(\mbox{b-abstraction}) of $x \in X$ if $\gamma.x^\sharp \slq x$. The triple
$\langle \Gamma, \Gamma^\sharp, \gamma \rangle$ is called
\emph{abstraction}, where $\Gamma$ is the \emph{concrete domain} and
$\Gamma^\sharp$ the \emph{abstract domain}.

\begin{definition} \label{def:b-abstraction_of}
Let $f : X \totype X$ be a monotone function, $f^\sharp
: X^\sharp \totype X^\sharp$ is a \emph{\mbox{b-abstraction}} of $f$ if
\[
( \forall x : X,\: x^\sharp : X^\sharp \;\;|\;\; 
  \gamma.x^\sharp \slq x 
  \;\:\dotsep\:\;
  \gamma.(f^\sharp.x^\sharp) \slq f.x
)\enspace.
\]
\end{definition}


\noi The next theorem, relevant to our work, shows that fixed points can be
calculated accurately and uniformly in an abstract domain.  
 
\begin{theorem} \label{thm:abst_fp}
  Let $\langle \Gamma, \Gamma^\sharp, \gamma \rangle$ be an abstraction 
  where $\Gamma$ is an almost complete \mbox{meet-semilattice} with a least
  element $\bot$ and $\Gamma^\sharp$ has a
  least element $\bot_\sharp$ with $\gamma.\bot_\sharp = \bot$.
  Let also $f : X \totype X$ be a monotone function and $f^\sharp : X^\sharp \totype X^\sharp$
  a \mbox{b-abstraction} of $f$. Then, if $f$ has a pre-fixed point in $X$
  \begin{enumerate}
  \item \label{it:abst_fp_is_lq}
    \remarkD{Estaria bueno probar $\gamma.(\sqcup \, i:\Nat \dotsep f^{\sharp (i)}.\bot_\sharp )
      \slq  \mu.f$}
    the supremum of $\{i:\Nat \dotsep \gamma.(f^{\sharp
    (i)}.\bot_\sharp)\}$ exists and is a lower bound of the least fixed point;
    that
    is $
    (\sqcup \, i:\Nat \dotsep \gamma.(f^{\sharp (i)}.\bot_\sharp) )
    \slq
    \mu.f \enspace,
    $

  \item \label{it:abst_fp_is_fp}
    \remarkD{Esto es realmente bueno ya que nos daria una forma de
      saber cuando el punto fijo es exacto.}
    \sloppy if 
    $(\sqcup \, i:\Nat \dotsep \gamma.(f^{\sharp(i)}.\bot_\sharp))$ is a
    pre-fixed point of $f$ then both definitions coincide; that
    is \mbox{$
    (\sqcup \, i:\Nat \dotsep \gamma.(f^{\sharp (i)}.\bot_\sharp))
    =
    \mu.f \enspace.
    $}
  \end{enumerate}

\end{theorem}
\begin{proof}\ \\*[-3ex]
  \begin{enumerate}

  \item
    First, we prove $\mu.f$ is an upper bound of the set \mbox{$\{i:\Nat \dotsep
      \gamma.(f^{\sharp(i)}.\bot_\sharp)\}$}. By induction: If $\gamma.\bot_\sharp =
    \bot$, then $\gamma.\bot_\sharp \slq \mu.f$\enspace. Suppose $\gamma.(f^{\sharp
      (i)}.\bot_\sharp) \slq \mu.f$ then
      \begin{align*}
        \gamma.(f^{\sharp (i+1)}.\bot_\sharp) 
            &= \gamma.(f^\sharp.(f^{\sharp (i)}.\bot_\sharp)) & \\
            &\slq f.(\mu.f) 
                 &\mbox{\{ inductive hypothesis, 
                           Definition \ref{def:b-abstraction_of} \}} \\
               &= \mu.f 
                 &\mbox{\{ $\mu.f$ is fixed point of $f$ \}}
      \end{align*}
    Hence, $\mu.f$ is an upper bound of that set.
    By Lemma~\ref{lem:priestley}, the
    supremum $(\sqcup \, i:\Nat \dotsep \gamma.(f^{\sharp (i)}.\bot_\sharp))$ exists and is lower
    than~$\mu.f$\enspace.

  \item 
    If $(\sqcup \, i:\Nat \dotsep \gamma.(f^{\sharp(i)}.\bot_\sharp))$ 
    is a \mbox{pre-fixed} point of $f$, by Theorem~\ref{th:fixpoint} 
    $
    \mu.f
    \slq
    (\sqcup \, i:\Nat \dotsep \gamma.(f^{\sharp (i)}.\bot_\sharp))
    $
    and by (\ref{it:abst_fp_is_lq}) these are equal.
  \end{enumerate}
\end{proof}

\noi The first part of this theorem shows that we can obtain a lower bound of
the least fixed point over an abstract domain. Then, given a program $P$ and an
abstract domain, this lower bound can be used to prove the correctness of the program
(\ref{eq:initialization_vc}). The second part gives us a sufficient condition to
achieve exact fix point calculation in the abstract domain. We will go into
these points in the next section.

\subsection{Random Variable Abstraction}
\label{subsec:rva}

By Theorem~\ref{thm:abst_fp} if we choose an appropriate abstract domain then we
could calculate or \mbox{under-approximate} the fixed point~$\mu.f$ (defined in
Section \ref{sec:correctness}) in order to prove the validity of~(\ref{eq:initialization_vc}). We begin by defining a suitable abstraction
$\langle \Gamma, \Gamma^\sharp, \gamma \rangle$. The basic idea consists in
generalizing predicate-based abstraction theory (PA)
\cite{graf97pa,uribe98abstractions} to expectations. In PA the abstraction is
determined by a set of $N$ linear predicates $\{p_1,\cdots, p_N\}$. The abstract
state space is just the lattice generated by the set of all \mbox{bit-vectors}
$(b_1,\cdots,b_N)$ of length $N$ whose concretization is defined as
$\gamma.(b_1,\cdots,b_N) = (\wedge i:[1..N] \dotsep b_i \Eq p_i)$. Then, any
predicate in the concrete domain can be represented as a series of truth values
over each of these atoms. Our RVA generalize these values by linear functions on
the program variables.
 
\noi Given a program $P = (S,\Init,C)$ over the set of variables
$X=\{x_1,\cdots,x_n\}$ and a set of disjoint convex linear predicates
$\phi_1,\cdots, \phi_m$ the abstract domain
will be $\Gamma^\sharp = \langle \Real^{(n+1)\times m}, \elq^\sharp
\rangle$. It can be thought as the set of parameters of linear
expectations over the program variables for each linear predicate
$\phi_i$. Hence, the concretization function is defined as:
\begin{equation} \label{eq:concr_func}
\gamma.((q_0^1,\cdots,q_n^1), \cdots, (q_0^m,\cdots,q_n^m))
\Def
\textstyle
(\sum i : [1..m] \;\dotsep\; 
  (q_0^i + (\sum j : [1..n] \dotsep q_j^i \times x_j) \times [\phi_i]))
\end{equation}
and the order relation as $x^\sharp \aelq y^\sharp \Def \gamma.x^\sharp \elq
\gamma.y^\sharp$\enspace. The predicates $\phi_1,\cdots, \phi_m$ can be
constructed from the guards of the program~\cite{graf97pa,uribe98abstractions}
and taking the complete boolean algebra~\cite{IntroductionLatticeOrderBook} over
this finite set of assertions.

\smallskip

The next step is to define a \mbox{b-abstraction} of the transformer
(\ref{eq:back_propag_def}). First note, as we want to approximate $\mu.f$ by
$(\sqcup \, i:\Nat \dotsep \gamma.(f^{\sharp (i)}.\bot_\sharp))$
(Theorem~\ref{thm:abst_fp}), we only need to obtain the set $\{i:\Nat \dotsep
f^{\sharp(i)}.\bot_\sharp\}$ such that expectations in $\{i:\Nat \dotsep
\gamma.(f^{\sharp(i)}.\bot_\sharp)\}$ form an ascending chain in $\ES$.
Thus, using Theorem~\ref{thm:abst_fp}, we can approximate by calculation the
least fixed point of the transformer as the limit of this chain. Moreover, the
function $f^\sharp : \Real^{(n+1)\times m} \totype \Real^{(n+1)\times m}$ over
the elements of the defined set must agree with
Definition~\ref{def:b-abstraction_of}, that is 
\[
( \forall x : \ES, i : \Nat \;\;|\;\; 
  \gamma.(f^{\sharp (i)}.\bot_\sharp) \elq x 
  \;\:\dotsep\:\;
  \gamma.(f^{\sharp (i+1)}.\bot_\sharp) \elq f.x
)\enspace.
\]
If we suppose
\begin{equation}\label{eq:sup_fabst}
\gamma.(f^{\sharp (i+1)}.\bot_\sharp) \elq f.(\gamma.(f^{\sharp(i)}.\bot_\sharp))
\end{equation}
then $f^\sharp$ agrees the above definition:
\begin{align*}
\gamma.(f^{\sharp (i)}.\bot_\sharp) \elq x 
  &\;\Impl\; f.(\gamma.(f^{\sharp (i)}.\bot_\sharp)) \elq f.x
    &\mbox{\{ $f$ monotone \}} 
\\&\;\Eq\; \gamma.(f^{\sharp (i+1)}.\bot_\sharp) \elq f.(\gamma.(f^{\sharp(i)}.\bot_\sharp)) 
  \\ &\qquad \And f.(\gamma.(f^{\sharp (i)}.\bot_\sharp)) \elq f.x
    &\mbox{\{ by supposition (\ref{eq:sup_fabst}) \}}
\\&\;\Impl\; \gamma.(f^{\sharp (i+1)}.\bot_\sharp) \elq f.x
    &\mbox{\{ by transitivity of ${}\elq{}$ \}}
\end{align*}


Taking into account (\ref{eq:sup_fabst}) we construct the set $\{i:\Nat \dotsep
f^{\sharp(i)}.\bot_\sharp\}$, where the bottom element~$\bot_\sharp$ is clearly the
null matrix in $\Real^{(n+1) \times m}$. Also, if we have defined
$f^{\sharp(i)}.\bot_\sharp$ we obtain $f^{\sharp(i+1)}.\bot_\sharp$ as
follows: first we calculate $f.(\gamma.(f^{\sharp(i)}.\bot_\sharp))$ and then we
choose an expectation in the lhs of (\ref{eq:sup_fabst}) that can be accurately
represented in $\Gamma^\sharp$. Let $f^{\sharp(i)}.\bot_\sharp =
((q_0^1,\cdots,q_n^1), \cdots, (q_0^m,\cdots,q_n^m))$ and $\beta^\sharp$ a
\mbox{b-abstraction} of $\beta$. 
The former can be obtained:
\begin{align*}
f.(\gamma.(f^{\sharp(i)}.\bot_\sharp)) 
  &= 
\textstyle
[G] \times
\Wp.P.
(\sum i : [1..m] \dotsep 
  (q_0^i + (\sum j : [1..n] \dotsep q_j^i \times x_j) \times [\phi_i]))
\\  &\quad + [\Not G] \times \gamma.\beta^\sharp
      & \mbox{\{ by (\ref{eq:concr_func}) and (\ref{eq:back_propag_def}) \}}
\\[1ex]
&=
\textstyle
[G] \times
( \sum I:\PowerNonEmpty[0..l) \dotsep
	[A_I] \times ( \bigsqcap r:I \dotsep
				( \sum s:[1..k_r] \dotsep p_r^s\times \eta_r^s)
			  )
)
\\ &\quad + [\Not G] \times \gamma.\beta^\sharp
      & \mbox{\{ by (\ref{eq:wp_sint}) \}}
\\
\text{where}
\\
&
\textstyle
\eta_r^s = 
(\sum i : [1..m] \dotsep 
  (q_0^i + (\sum j : [1..n] \dotsep q_j^i \times x_j)\Subst{E_r^s} \times [\phi_i\Subst{E_r^s}]))
\enspace.
\end{align*}

\noi Then, we restrict the above result within each domain defined by
predicates $\phi_i$ obtaining a set of expectations.
Although, these expectations are not necessarily linear, if the program is
linear (we have only linear assignments $E_r^s$), each of them can be
lower bounded by linear functions $g_i$ agreeing~(\ref{eq:sup_fabst}). Also,
we choose each function $g_i$ such that they are greater or equal than $[\phi_i]
\times\gamma.(f^{\sharp (i)}.\bot_\sharp)$ ensuring monotonicity of the
constructed chain. So, each $g_i$ must be bounded above and below:
\[
[\phi_i] \times \gamma.(f^{\sharp (i)}.\bot_\sharp) 
\;\elq\;
[\phi_i] \times g_i
\;\elq\; 
[\phi_i] \times f.(\gamma.(f^{\sharp(i)}.\bot_\sharp))\enspace.
\]
\sloppy Note that it is always possible to find these bounded expectations $g_i$ because
$f$ is monotone (then $\gamma.(f^{\sharp (i)}.\bot_\sharp) \elq
f.(\gamma.(f^{\sharp(i)}.\bot_\sharp))$) and the lower bound $[\phi_i] \times
\gamma.(f^{\sharp (i)}.\bot_\sharp)$ defines an hyperplane over the set of
states $\phi_i$. Also, as $g_1,\cdots,g_m$ are linear functions, we can find the parameters
\mbox{$(({q'}_0^1,\cdots,{q'}_n^1), \cdots, ({q'}_0^m,\cdots,{q'}_n^m)) \in
\Real^{(n+1)\times m}$} such that each $g_i.(x_1,\cdots,x_n) \Def {q'}_0^i + (\sum j
: [1..n] \dotsep {q'}_j^i \times x_j)$. Then we define
$f^{\sharp(i+1)}.\bot_\sharp \Def (({q'}_0^1,\cdots,{q'}_n^1), \cdots,
({q'}_0^m,\cdots,{q'}_n^m))$. Thus, we abstract
$f.(\gamma.(f^{\sharp(i)}.\bot_\sharp))$ by choosing a lower expectation that
could be accurately represented in $\Gamma^\sharp$.

\noi Henceforth, by applying this method we can reach a fixed point
$\varphi^\sharp$ in the abstract domain provided that exists a pre-fixed point
of transformer $f$ (condition from Theorem~\ref{thm:abst_fp}). By
Theorem~\ref{thm:abst_fp}(\ref{it:abst_fp_is_lq}), expectation
$\gamma.\varphi^\sharp$ is a lower bound of $\mu.f$ and can be used to prove the
correctness of the program regarding \mbox{post-expectation}~$\beta$. Also, if we can
check $\gamma.\varphi^\sharp$ is a pre-fixed point of $f$ then, by
Theorem~\ref{thm:abst_fp}(\ref{it:abst_fp_is_fp}), it is the least fixed point of
$f$ and our abstraction works accurately.

The following section shows an example using this method.

\section{Martingale Example}\label{case_study}

In this section we give a more detailed example of our probabilistic program verification through random variable abstraction, computing the average capital a gambler would expect from a typical martingale betting system (Fig.~\ref{fig:martingale}).
The gambler starts with a capital $C$ and bets initially one unit.
While the bet $b$ is lower than the remaining capital $c$, a fair coin is tossed and if it is a win, the player obtain one hundred percent gain and stops gambling, otherwise it doubles the bet and continues.
\begin{figure}[!ht]
\begin{center}
\begin{minipage}{\textwidth}
\begin{tabbing}
	$\Prg_1$\ $\Def$ \= \ \= $c,b:=C,1$ \\
	\>; \> $\DO$ \= $0<b\leq c \ra$ \\
	\>\>	\> \ \= $c:=c-b$ \\
	\>\>	\> ; \> $c,b:=c+2b,0\ \oplus_{\frac{1}{2}}\ b:=2b$ \\
	\>\> $\OD$
\end{tabbing}
\end{minipage}
\end{center}
\caption{Martingale.}
\label{fig:martingale}
\end{figure}
The expected gain of this strategy is given by $\Wp.\Prg_1.c$ .

\begin{figure}[!ht]
\begin{center}
\includegraphics[trim=0 1.1cm 0 1.1cm, clip, scale=0.75]{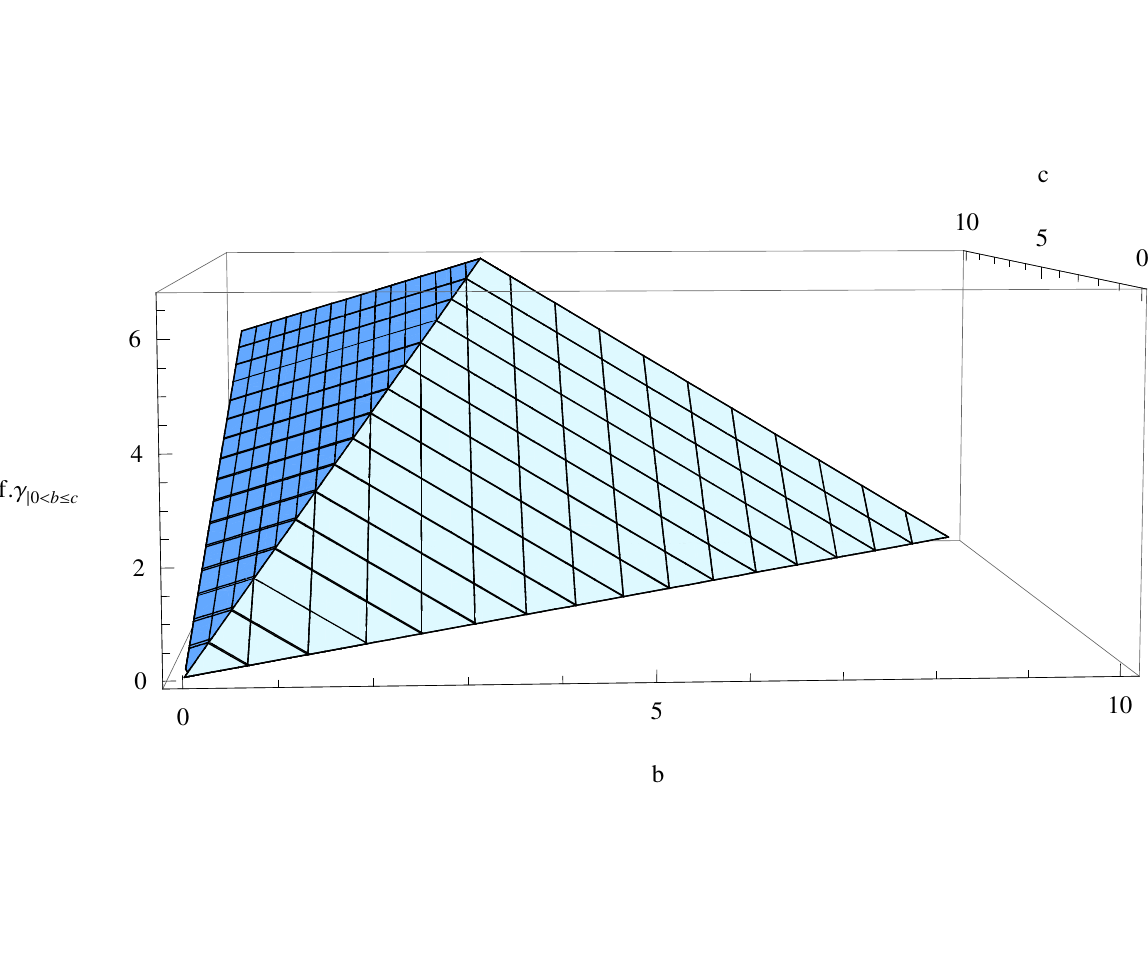}
\caption{Piecewise linear function to be minimized by a plane.}
\label{fig:cuna}
\end{center}
\end{figure}

Following Section \ref{subsec:rva} we first define the set of variables $X$ and the set of disjoint convex linear predicates $\phi_i$.
For the variables we take the whole set $\{b,c\}$ since both participate in the control flow.
The predicates involved should, in principle, include the guard $0<b\leq c$ and its negation as it is customary in PA.
However the update operations in the program produce movements in the bidimensional state space that cannot be accurately captured by this two regions.
We propose an abstraction that divides the two dimensional space $(b,c)$ into six different regions (predicates).
The regions are defined by the four atoms given by the inequalities $0<b$, $b\leq c$, plus two other inequalities that define the six total orders between $0,b$ and $c$.
The concretization function $\gamma$ of the abstract domain $\Real^{(2+1)\times 6}$ is the sum of a linear function on $b,c$ in each region.
%
\begin{align*}
\phi_1 &= 0 \leq c < b  &  \phi_4 &= b \leq c < 0 & \gamma \Def \textstyle \sum_{i=1}^{6} (c_i c + b_i b + a_i) [\phi_i] \\
\phi_2 &= 0 < b \leq c  &  \phi_5 &= c < b \leq 0 \\
\phi_3 &= b \leq 0 \leq c  &  \phi_6 &= c < 0 < b
\end{align*}
If we disregard the initialization and coalesce the two assignments, the program can be transformed into an equivalent $\Prg_1'$, but following the syntax of Section~\ref{concrete_semantics}.
It consists of just one guard:
\[ \Prg_1' \Def 0<b\leq c \ra (c,b:=c+b,0)@\textstyle{\frac{1}{2}} \mid (c,b:=c-b,2b)@\textstyle{\frac{1}{2}}\enspace. \]
The fixed point computation is $f.X \Def [G]\times\Wp.\Prg_1'.X + [\Not G]\times\beta$ with \mbox{post-expectation} $\beta=c$.
The term $\Wp.\Prg_1'.X$ is given by (\ref{eq:wp_sint}) and the other summand is given by the final state condition on the \mbox{post-expectation}.
\[ f.X = [0<b\leq c] \times (\tfrac{1}{2} X\Subst{c,b:=c+b,0} + \tfrac{1}{2}
X\Subst{c,b:=c-b,2b}) +  [\Not(0<b\leq c)]\times c \enspace. \]
The function $f^\sharp$ is b-abstraction of $f$ that we are going to calculate now.
We apply $f$ to the concretization function $\gamma$ over generic coefficients, and obtain three blocks of summands given by the left assignment, the right assignment and the \mbox{post-expectation}.
%
\begin{align*}
	f.\gamma &= \tfrac{1}{2}(c_3 c + c_3 b + a_3)[0<c+b] \\
	&\quad {}+ \tfrac{1}{2}\left( (c_1 c+(2b_1-c_1)b+a_1)[b\leq c<3b] + (c_2 c+(2b_2-c_2)b+a_2)[0<2b \land 3b\leq c] \right) \\
	&\quad {}+ (\textstyle{\sum} i:[1..6] \mid i\neq 2 \dotsep (1c+0b+0) [\phi_i])\enspace.
\end{align*}
%
Clearly the first three summands do not fit into the abstraction given by predicates $\phi_i$,
however we can give linear lower bounds on each of them that can be accurately represented in the abstract domain $\Gamma^\sharp$ as stated in Section \ref{subsec:rva}.
The region $0\leq c+b$ includes $\phi_2$ or in terms of random variables $[0<b\leq c] \elq [0\leq c+b]$, so we keep the linear function in the smaller region.
The other two regions $\Psi_1 \Def b\leq c<3b$, $\Psi_2 \Def 0<3b\leq c$ form a partition of $\phi_2$.
It is needed to find a plane that in $\Psi_1$ is below $(c_1 c+(2b_1-c_1)b+a_1)$ and in $\Psi_2$ is below $(c_2 c+(2b_2-c_2)b+a_2)$.
%
%
The two planes form a wedge (Fig.~\ref{fig:cuna}), so our lower bounding plane would put a base on it.
In the region $\Psi_1$ the function simplifies to $c-b$ and the minimum in the region is $0$ when $c=b$.
For $\Psi_2$ the minimum is $c_2$ achieved in $b=0$.
Putting it all together we obtain the final linear expression that bounds $f.\gamma$ from below and it is in terms of the original predicates:
\[ f.\gamma \geq (\tfrac{c_2+1}{2} c + \tfrac{1-c_2}{2} b)[\phi_2] + (\textstyle{\sum} i:[1..6] \mid i\neq 2 \dotsep (1c+0b+0)[\phi_i]) \enspace.\]
Note that doing this we comply with Eq. \ref{eq:sup_fabst}, and this in turn implies that $f^\sharp$ is a \mbox{b-abstraction}.
We explicitly write down $f^\sharp$ and iterate it up-to fixed point convergence in Table~\ref{table:fp_martingale}.
As expected, $\phi_4,\phi_5,\phi_6$ do not contribute in the fixed point computation since $0\leq c$ is a program invariant, so we do not include them:
\[ f^\sharp.((c_1,b_1,a_1),(c_2,b_2,a_2),(c_3,b_3,a_3)) \Def
	((1,0,0),(\tfrac{c_2+1}{2},\tfrac{1-c_2}{2},0),(1,0,0)) \enspace.
\]
\begin{table}[!ht]
\begin{small}
\begin{center}
\begin{tabular}{|c||c|l|c|}
\hline
& \multicolumn{3}{|c|}{Coefficients} \\
\hline
Iter. & ($c_1$, $b_1$, $a_1$) & ($c_2$, $b_2$, $a_2$) & ($c_3$, $b_3$, $a_3$)  \\
\hline
0  &  (0, 0, 0) & (0, 0, 0) & (0, 0, 0) \\
1  &  (1, 0, 0) & (0.5, 0.5, 0) & (1, 0, 0) \\
2  &  (1, 0, 0) & (0.75, 0.25, 0) & (1, 0, 0) \\
3  &  (1, 0, 0) & (0.875, 0.125, 0) & (1, 0, 0) \\
4  &  (1, 0, 0) & (0.9375, 0.0625, 0) & (1, 0, 0) \\
5  &  (1, 0, 0) & (0.96875, 0.03125, 0) & (1, 0, 0) \\
6  &  (1, 0, 0) & (0.984375, 0.015625, 0) & (1, 0, 0) \\
7  &  (1, 0, 0) & (0.9921875, 0.0078125, 0) & (1, 0, 0) \\
\vdots & \vdots & \phantom{(1,}\hspace{0.9ex}\vdots & \vdots  \\
55  &  (1, 0, 0) & (1, 0, 0)  & (1, 0, 0) \\
\hline
\end{tabular}
\end{center}
\end{small}
\caption{Iteration for the martingale.}
\label{table:fp_martingale}
\end{table}

%
%
%

%
%

We can see $f^{\sharp (i)}.\bot_\sharp \aelq f^{\sharp (i+1)}.\bot_\sharp$, so the under approximation $\gamma.(f^{\sharp (i+1)}.\bot_\sharp)$ tends from below to $(\textstyle{\sum} i:[1..6] \dotsep (1c+0b+0)[\phi_i]) = c$.

The initialization $c,b:=C,1$ does the rest obtaining a lower bound on the expected remaining capital, that is precisely $C$.
It can be proven that $c$ is a prefix point of $f$ (in fact it is a fixed point),
and using Theorem~\ref{thm:abst_fp}~(\ref{it:abst_fp_is_fp}), we establish $\mu.f = c$, therefore our approximation is exact.
Moreover, as the program is purely probabilistic, $C$ is the expectation of $c$ with respect to $P$.

\section{Aims and conclusions}\label{conclusions}

This paper presents a technique for computing the expectation of unbounded random
variables tailored to performance measures, like expected number of rounds of a
probabilistic algorithm or expected number of detections in anonymity protocols.
Our method can check quantitative linear properties on any denumerable state space using
iterative backwards fixed point calculation.


Perhaps the main drawback of the method is that it is semi-computable but it covers cases
where previous work cannot be applied (geometric distribution,
martingales). Besides, it seems hard to bound expectations of programs syntactically since a
minor (linear) modification in the geometric distribution algorithm leads to a
unbounded expectation for a program that halts with probability $1$.

\bigskip

In future work we would like to build tool support for our approach. This would
involve, among other tasks, the mechanization of the weakest pre-expectation
calculus in the abstract domain, as well as the maximization problem that involves computing a lower linear function in each iteration.
As our technique works on linear domains, this later task would be easily solved by known linear programming techniques.

We also plan to analyze more complex programs.
The Crowds anonymity protocol modeled as in~\cite{shmatikov04crowds} is a good candidate for our automatic quantitative program analysis, since its essential anonymity properties are expressed as the expected number of times the message initiator is observed by the adversary with respect to the observations obtained for the rest of the crowd.
It is also planned to reproduce the results of Rabin and Lehmann's probabilistic dining-philosophers algorithm~\cite{mciver02expected}.


\smallskip

\textit{Acknowledgements.} We would like to thank Pedro D'Argenio for his support in this project, as well as Joost-Pieter Katoen for handing us an early draft of~\cite{katoen09linear}.

\bibliographystyle{eptcs}
\bibliography{bibliography}

\end{document}